\documentclass[journal]{IEEEtran}
\usepackage[caption=false]{subfig}
\usepackage{graphicx}
\usepackage{tabularx}
\usepackage{amsmath,amssymb,bm,mathrsfs,xfrac,mathtools}
\usepackage{color}
\usepackage[export]{adjustbox}
\usepackage{float}
\usepackage{cite}
\usepackage{tablefootnote}
\renewcommand{\baselinestretch}{1} % stretch line spacing

\usepackage{standalone}
\usepackage{tikz}
\usetikzlibrary{calc,shapes,arrows,automata,positioning}
\definecolor{gray45}{rgb}{0.2, 0.5, 0.478}
\definecolor{ffqqqq}{rgb}{1,0,0}
\definecolor{qqzzff}{rgb}{0,0.6,1}
\tikzstyle{startstop} = [rectangle, rounded corners, minimum width=0.5cm, minimum height=0.8cm,text centered, draw=black, fill=green!10]
\tikzstyle{io} = [trapezium, trapezium left angle=70, trapezium right angle=110, minimum width=0.5cm, minimum height=0.8cm, text centered, draw=black, fill=blue!30]
\tikzstyle{process} = [rectangle, minimum width=0.5cm, minimum height=0.8cm, text centered, draw=black, fill=white!100]
\tikzstyle{decision} = [diamond, minimum width=0.5cm, minimum height=0.8cm, text centered, draw=black, fill=yellow!20]
\tikzstyle{arrow} = [thick,->,>=stealth]

\ifCLASSINFOpdf
\else
\fi

\begin{document}

\title{Roles~of~Dynamic~State~Estimation~in~Power~System Modeling, Monitoring and Operation}

%\author{\emph{IEEE Task Force on Power System Dynamic State and Parameter Estimation}\\
%Contributing Authors: \textbf{(a complete list will be determined later)}
%}
\author{IEEE Task Force on Power System Dynamic State and Parameter Estimation\\
	Junbo Zhao (TF Chair), \IEEEmembership{Senior Member,~IEEE}, Marcos Netto, \IEEEmembership{Member,~IEEE}, Zhenyu Huang, \IEEEmembership{Fellow,~IEEE}, Samson Shenglong Yu, \IEEEmembership{Member,~IEEE}, Antonio~G\'{o}mez-Exp\'{o}sito, \IEEEmembership{Fellow,~IEEE}, Shaobu Wang, \IEEEmembership{Senior Member,~IEEE}, Innocent Kamwa, \IEEEmembership{Fellow,~IEEE}, Shahrokh Akhlaghi, \IEEEmembership{Member,~IEEE}, Lamine Mili, \IEEEmembership{Life Fellow,~IEEE}, Vladimir Terzija, \IEEEmembership{Fellow,~IEEE}, A. P. Sakis Meliopoulos, \IEEEmembership{Fellow,~IEEE}, Bikash Pal, \IEEEmembership{Fellow,~IEEE}, Abhinav Kumar Singh,~\IEEEmembership{Member,~IEEE}, Ali Abur, \IEEEmembership{Fellow,~IEEE}, Tianshu Bi, \IEEEmembership{Senior Member,~IEEE}, Alireza Rouhani, \IEEEmembership{Member,~IEEE}
}

\markboth{IEEE TRANSACTIONS ON POWER SYSTEMS,~Vol.~, No.~, ~2020}
{Shell \MakeLowercase{\textit{et al.}}: Bare Demo of IEEEtran.cls for Journals}

\maketitle

\begin{abstract}
Power system dynamic state estimation (DSE) remains an active research area. This is driven by the absence of accurate models, the increasing availability of fast-sampled, time-synchronized measurements, and the advances in the capability, scalability, and affordability of computing and communications. This paper discusses the advantages of DSE as compared to static state estimation, and the implementation differences between the two, including the measurement configuration, modeling framework and support software features. The important roles of DSE are discussed from modeling, monitoring and operation aspects for today's synchronous machine dominated systems and the future power electronics-interfaced generation systems. Several examples are presented to demonstrate the benefits of DSE on enhancing the operational robustness and resilience of 21st century power system through time critical applications. Future research directions are identified and discussed, paving the way for developing the next generation of energy management systems.
\end{abstract}


\IEEEpeerreviewmaketitle

\section{Introduction}
\IEEEPARstart{D}{ynamic} state estimation (DSE) \cite{Ali_chapter2016} is going to be very useful for time critical monitoring, control, and protection of future electric power grids. This is largely due to the changes in generation mixes and load compositions, particularly the increasing penetration of intermittent, stochastic and power electronics-interfaced non-synchronous renewable generation and distributed energy resources (DERs) \cite{Ben2017}. In this context, where the system operating point changes more often and more rapidly, tracking the system dynamic state variables is of critical importance. Recall that dynamic state variables are the ones associated with the time derivatives in the set of differential-algebraic equations describing power system dynamics \cite{Peter2019}. The dynamic state variables of the synchronous generators correspond to their electromagnetic and electromechanical processes, as well as their controllers \cite{Junbo_TF2019}. As for the non-synchronous generations, the dynamic state variables are associated with the primary source of energy, e.g. solar photovoltaic arrays, batteries and wind turbines, and their controllers. Note that the phenomena driving the primary source of energy in this case are diverse. Examples include frequency at the point of common coupling between the power converter and the electric grid, the electric current in the converter, and the pitch angle of wind turbines. Furthermore, 
there exists an important challenge in the development and maintenance of accurate models for power electronics-interfaced devices. DSE can be developed to validate the models and to estimate unknown or incorrect parameters.

Despite being initially mentioned in the 1970s \cite{Miller1971}, it was only in recent years that the power system community has picked up the momentum in DSE research. Part of the reason was the lack of appropriate metering infrastructure, like phasor measurement units (PMUs) and merging unit (MUs) that are being widely deployed to capture the appropriate dynamics in power systems \cite{Phadke2010_PMU,Song_MUs_2017}. Feasibility studies using PMU measurements for DSE are reported in \cite{Henry2007}, and subsequently, various Kalman filtering techniques, such as Extended Kalman Filter (EKF), Unscented Kalman Filter (UKF) \cite{Kamwa2011}, Ensemble Kalman Filter (EnKF), Particle Filter (PF) and their variants \cite{NingZhou2015} have been applied to DSE. Data-driven DSE \cite{Marcos2018}, observability analysis to guide measurement selection \cite{Abur_observability2017} and the enhancement of robustness against bad data and parameter errors are also developed in \cite{Junbo_GMIEKF2017,Junbo_GMUKF2019,Abur2018}. Readers may refer to \cite{Junbo_TF2019} for a comprehensive summary of DSE algorithms.

Several online tasks can benefit from DSE but have not adequately deliberated the need for it. An example is the dynamic security assessment (DSA). Today’s DSA tool assumes a steady-state initial condition \cite{Leiwang_DSA_2006}, which would yield inaccurate results in the presence of more fluctuations from DERs and loads. While DSE could directly estimate the dynamic states and therefore provide more accurate initialization conditions \cite{Henry_DSA_2014}. Furthermore, for synchronous generators, numerous developed control schemes rely on power system stabilizers (PSS) to damp out oscillations. These control schemes use the rotor’s speed as input. The direct use of measurements obtained from a meter installed on the shaft of the machine is not reliable; instead, the current practice is to use the compensated frequency \cite{IEEE_PSS_2016}, which is calculated using the voltage and current measured at the generators’ terminal. However, the calculation of the compensated frequency is significantly affected under transient conditions \cite{Su_PSS_2013}, which leads to sub-optimal control response of the PSS. If DSE is deployed for frequency estimation, this issue can be effectively resolved. The rate of change of frequency (ROCoF) estimation can also be formulated in the state- space model for DSE \cite{Abhinav2019}. Note that, in these examples, traditional static state estimation (SSE) could not provide reliable system dynamic states and quantities. Therefore, it is of critical importance to understand the role of DSE in power systems from various aspects. While SSE has been in energy management systems (EMS) for decades and its roles and implementation requirements are well established, the same does not apply for DSE. There is a large gap in understanding different roles of SSE and DSE and how DSE can be implemented in practice. 

This paper summarizes the joint efforts of the Task Force in clarifying the roles of DSE from the power system modeling, monitoring, and operation perspectives. Comparisons between SSE and DSE in practical implementations are thoroughly discussed, including measurement, model requirements, software support and potential applications based on the estimation results. Illustrative examples and summary frameworks have been presented to appreciate the roles of DSE. The future applications are discussed to shed light on the transition from today’s EMS to its next generation. 

The remainder of this paper is organized as follows. Section II compares DSE and SSE from the practical implementation requirement and application aspects. Section III summarizes the role of DSE in different applications. Conclusions and directions for future research are given in Section IV.

\begin{table}[htb!] 
\caption{Comparison Between SSE and DSE.}
\centering \footnotesize
\setlength{\tabcolsep}{0.7em}
\begin{tabular}{l c c}
\hline\hline
 & SSE & DSE \\ \hline
Measurements & From the SCADA & From PMU/DFR \\
 & (every $\sim$2-10 sec.) & (every 1/30$\sim$1/240 sec.) \\ \hline
Observability & Binary\tablefootnote{Topological observability analysis provides a binary answer as the first byproduct, but it gives additional valuable information, such as which buses (islands) are observable or which pseudo-measurements should minimally be added to restore full observability. Numerical observability analysis, being based on the factorization of the Jacobian or Gain matrices, provides a ``spectrum" or range of observability answers, depending on the condition number of the matrix being factorized. A network can be topologically observable but not algebraically (numerically), owing to abnormal network parameters or measurement weights. In some cases, a network can be numerically observable for a certain combination of weights, but not for others. This is not exactly ``binary".} & Time-varying \\
 & (observable or not) & (Strong/weak/not observable) \\ \hline
Update speed & 1 snapshot & 1 prediction + 1 filtering \\
 & (every $\sim$2-10 sec.) & step of the Kalman filter \\
 & & (every $<$ 1/30 sec.) \\ \hline
Models & Algebraic power & Differential-algebraic \\
 & flow equations & equations \\ \hline
Framework & Mostly centralized & Centralized \& \\
 & or distributed & distributed/decentralized \\ \hline
Outputs & Algebraic variables & Dynamic variables \\
 & (voltage magnitudes & (machine/ dynamic load/ \\
 & and angles) & DERs dynamic variables) \\ \hline
Applications & Monitoring and control & Monitoring (operator in \\
 & (operator in the loop) & the loop), control and \\
 & & adaptive protection \\
 & & (operator out of the loop) \\
\hline\hline
\end{tabular}
\label{table2a} 
\end{table}

\vspace{-0.3cm}
\section{Comparative Overview of SSE and DSE}
SSE has become a widely used tool in today's EMS, while DSE is a new tool for the industry and system operators. It is essential to clarify their implementation and functionality differences and, at the same time, enable a clear path transition from SSE-based EMS to the future DSE-based EMS with power electronics-dominated power systems.

\vspace{-0.4cm}
\subsection{Implementation Differences}
SSE and DSE have different requirements with respect to measurements, models, observability, execution rate, outputs and applications. These differences are summarized in Table I. SSE mostly relies on SCADA measurements that are updated every 2-5s and some PMU measurements to gain more redundancy. But since SCADA measurements are not synchronized while those of PMUs are, effectively integrating those two sources of data needs to be taken care of. By contrast, for DSE, fast and synchronized measurements with a reporting rate of 30 to 240 samples per second, are used and they might come from PMUs and digital fault recorders (DFRs). Furthermore, there is a significant difference in the observability theory for SSE and DSE \cite{Abur_observability2017,Junjian_observability2015}. For SSE, the topological- or numerical-based observability analysis typically determines whether the system is observable or not. If the system is unobservable, observable islands may be determined \cite{Abur_book_2004}. Furthermore, in the presence of ampere measurements, where multiple solutions may exist depending on the nature of the remaining measurements and the loading point, the observability analysis can also inform whether there are multiple solutions or not to the SSE problem. A network can be uniquely observable for heavily loaded cases, but ``unobservable" (i.e., undefined Jacobian) when the ampere measurement value is nearly zero (unloaded line), which leads to 0/0 indetermination. Thus, the answer depends on the line loading but is typically binary. By contrast, in DSE, one may refer to strongly or weakly or not observable systems. One way to quantify this is to compute the smallest singular value of the observability matrix from the Lie-derivatives. Higher (lower) values of the smallest singular value of the observability matrix indicates stronger (weaker) observability for a given measurement set \cite{Abur_observability2017}. Furthermore, due to the nonlinear and time-dependent nature of the problem, observability results are time-varying. It is worth noting that under certain conditions, the system might be unobservable, but still detectable \cite{Ning_2020} for DSE. Detectability is a slightly weaker notion than observability. A system is detectable if all the unobservable states are stable. Also, each prediction-correction step of the DSE must be numerically solved faster than the PMUs/DFRs/MUs scan rate, thereby posing some challenges on the computational power. To address this issue, decentralized/distributed DSE or parallel computing technique for centralized DSE is usually suggested \cite{Junbo_TF2019}.  The outcome of SSE and DSE are also different. SSE provides estimates of the bus voltage magnitudes and phase angles, which are the algebraic variables; on the other hand, DSE provides estimates of the dynamic state variables, such as those associated with generators/dynamic loads/DERs. There is also joint DSE that estimates dynamic and algebraic variables simultaneously \cite{Junbo_TF2019}. For some PMU observable networks, linear state estimation (LSE) that keeps up with the PMU refreshing rate is developed \cite{Abur_LAV2014}. However, it does not track the actual system dynamics.

%\begin{figure}\centering
%\includestandalone[width=\linewidth]{tikz2}
%\caption{(a) Simplified node-breaker model of a typical power generation substation. (b) Bus-branch model of the static state estimation. (c) Bus-branch model of the dynamic-state estimation.}
%\label{f007}
%\end{figure}

When implementing SSE and DSE, the models used are different. In SSE, the generators and loads are simply modeled by power injections and, hence, the system model used for SSE is represented using algebraic equations. While for DSE, the generators/dynamic loads/DERs, etc., and their controllers are represented by a set of DAEs. Note that it is not required to have PMUs installed at each generator terminal; if the generator terminal is observable via a local LSE, the DSE can be implemented.

\begin{figure}[htb!]
\centering
\includegraphics[width=\linewidth]{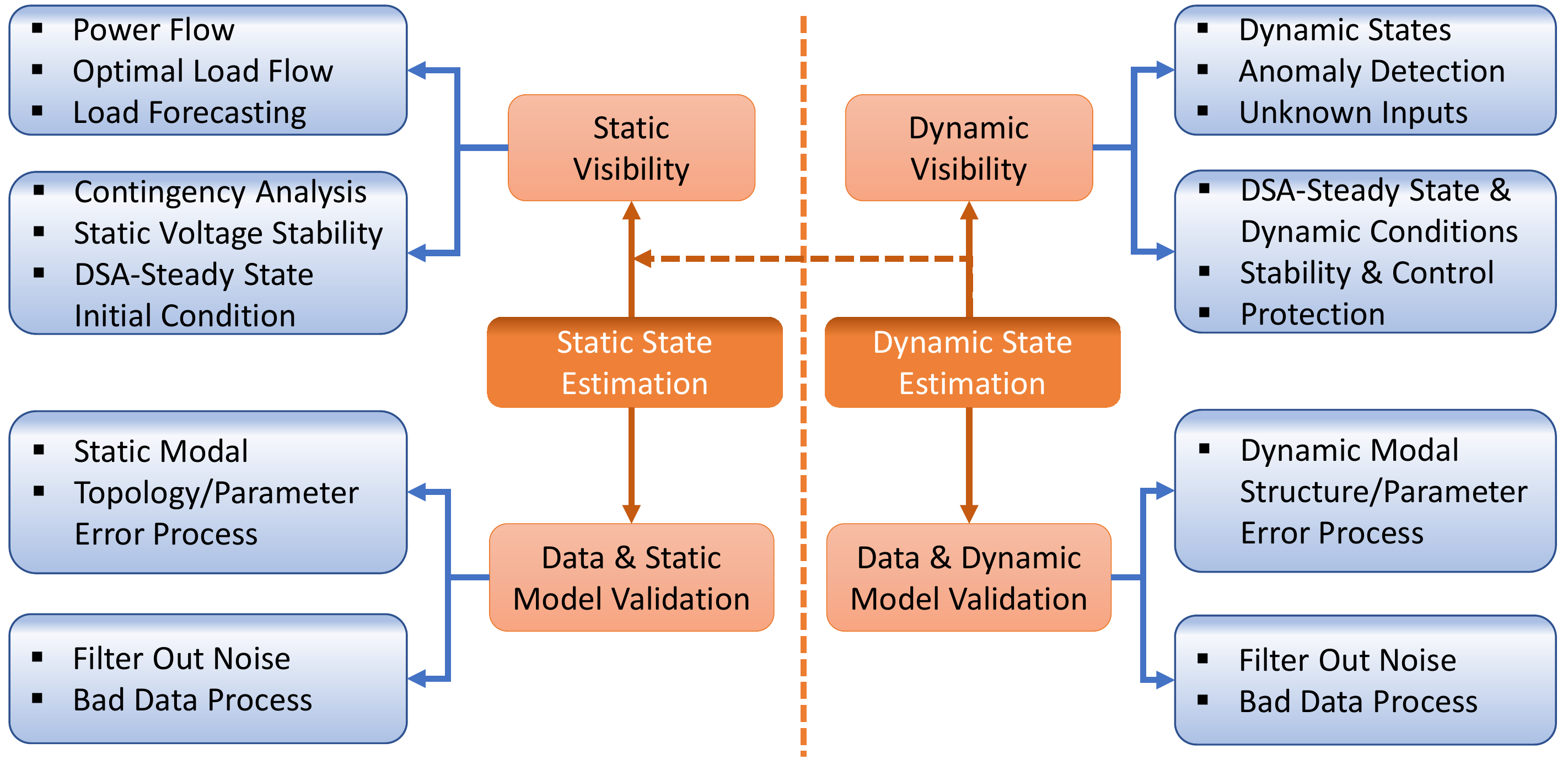}
\caption{Functional schematic comparisons between SSE and DSE.}
\label{fig2}
\end{figure}

\vspace{-0.3cm}
\subsection{Functionality Differences}
From the functionality perspective, both SSE and DSE have two essential roles, namely improving the targeted system visibility and validating the used data and models, which are summarized in Fig. \ref{fig2}. Since SSE is the snapshot-based static view of the system algebraic variables, its outputs allow the calculations of power injections and power flows across the systems. This provides the database for many applications in today’s EMS, including system operations, such as power flow, optimal power flow, load forecasting, etc., as well as security assessment, i.e., static contingency and static voltage stability analysis. By contrast, DSE relies on fast sampled and time synchronized measurements to track system dynamic changes, thus providing dynamic visibility, such as dynamic states, anomaly detection, unknown inputs estimation, etc. \cite{Abur_excitor_2016,Anagnostou2018,Wei_rotorstab_2018}. This dynamic information further allows other related applications, i.e., DSA \cite{Henry_DSA_2014}, rotor angle stability assessment \cite{Wei_rotorstab_2018}, adaptive power system protection \cite{Sakis_2017_protection} and so on. In terms of data and model validation for SSE, it relies on the measurement residual-based statistical test or robust estimation criteria for bad data processing and measurement model parameter error identifications. The bad data are caused by gross measurement errors or data transfer errors while model parameter errors can be due to transmission line parameters and topology inaccuracies. Since DSE relies on a set of DAEs, besides those parameter errors previously mentioned, it also encounters in equations that govern the system dynamics, e.g., generator models and their related controllers. These errors are classified into two categories: innovation outliers, which are caused by parameter and/or input errors in the set of DAEs; and structural outliers, which are caused by a model structure that does not reflect well the system dynamics \cite{Junbo_GMUKF_2018_TSP}. Note that the DSE outputs can be used to calculate the system algebraic variables as well. Therefore, the SSE functions can benefit from DSE. And vice versa, the information provided by the SSE can be used to warm start the DSE process before an event occurs.

\textbf{\emph{Remark:}} with the increasing deployment of PMUs, many high voltage transmission systems are already observable with only PMU measurements, such as that of the Dominion Energy, the 765/345/230 kV power grid in New York, and the 345 kV power grid in New England. As a result, the legacy nonlinear SSE algorithms can now be reformulated as the LSE \cite{Abur_LAV2014}. As long as the computational power is sufficient, the LSE can be updated with the scan rate of PMUs but its results are also restricted to algebraic variables-based applications. Compared with the SCADA-based SSE, an LSE allows for faster real-time contingency analysis and voltage stability assessment, besides area angle limit monitoring. By comparing LSE with DSE, LSE does not touch the dynamic equations and can not provide the real-time picture of system dynamic states and the related controllers. For example, the rotor speed information is not available from LSE and thus many applications based on them cannot be done. The importance of dynamic states for power system monitoring, visibility and operation will be highlighted in Section III. In summary, the fundamental differences in terms of potential applications are that LSE only deals with algebraic variables while DSE provides both dynamic and algebraic variables of the system. 

\vspace{-0.3cm}
\subsection{Practical Implementation of DSE}
\begin{figure}[htb!]
\centering
\includegraphics[width=\linewidth]{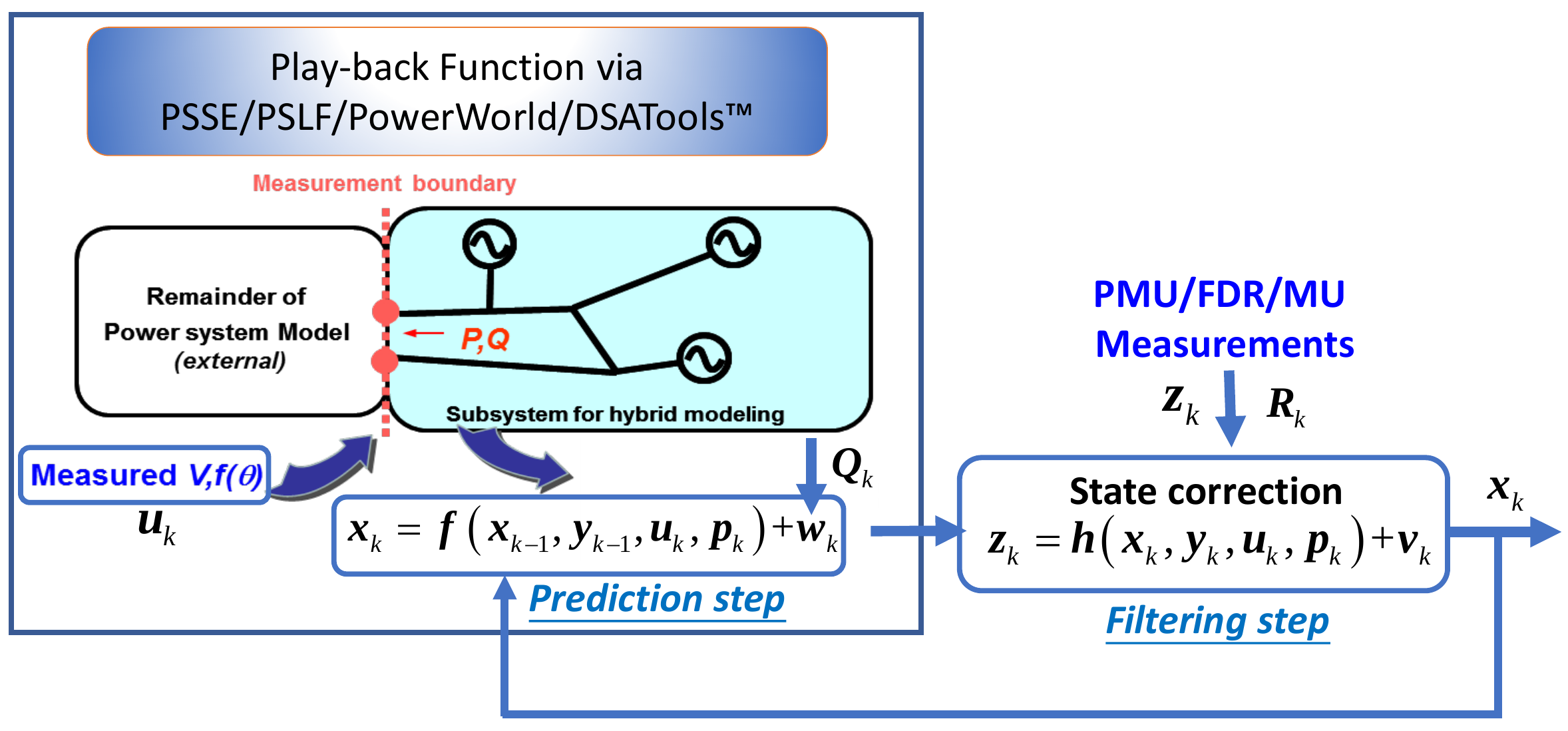}
\caption{Implementation of DSE using existing commercial tools.}
\label{fig3}
\end{figure}

There are several vendors who offer commercial software for SSE implementation while there is no commercial software for DSE. However, we can leverage the capabilities of existing commercial tools and enable the DSE implementation. The practical implementation of DSE is shown in Fig. \ref{fig3}, where the system is modeled by the following state-space model after time discretization of DAEs, and considering the available measurements:
\begin{align}
\bm{x}_{k} &= \bm{f}(\bm{x}_{k-1},\bm{y}_{k-1},\bm{u}_{k},\bm{p}_{k}) + \bm{w}_{k}, \; \mathbb{E}\left[\bm{w}_{k}\bm{w}_{k}^{\top}\right] = \bm{Q}_{k} \\
\bm{z}_{k} &= \bm{h}(\bm{x}_{k},\bm{y}_{k},\bm{u}_{k},\bm{p}_{k}) + \bm{v}_{k}, \; \mathbb{E}\left[\bm{v}_{k}\bm{v}_{k}^{\top}\right] = \bm{R}_{k} 
\end{align}
%\noindent
where $\bm{x}_{k}$ and $\bm{y}_{k}$ represent system dynamic and algebraic state vectors, respectively; $\bm{z}_{k}$ is the measurement vector from PMUs or DFR or MU; $\bm{u}_{k}$ is the system input vector that drives the state transition; $\bm{p}_{k}$ denotes the system parameters; $\bm{f}$ and $\bm{h}$ are vector-valued nonlinear functions; $\bm{w}_{k}$ and $\bm{v}_{k}$ are the system process and measurement error vector, respectively, with covariance matrices $\bm{Q}_{k}$ and $\bm{R}_{k}$.

The implementation of DSE has two main steps, namely the state prediction via (1) to obtain predicted state at time $k$ and the state correction that integrates the state prediction and measurements in (2) for filtering. It is interesting to note that the widely used commercial software, such as PSSE/ PSLF/ PowerWorld/ DSATools have the so-called event playback functions available. This play back function allows us to implement equations (1)-(2) automatically. The remaining work is to rely on Kalman filter framework and integrate the state prediction and the measurements together for the estimation of dynamic states. To fully leverage this playback function, it is suggested to use nonlinear Kalman filter techniques that do not require linearization, such as UKF, EnKF, PF and their variants. Another useful implementation of DSE is to take unknown dynamic states or control inputs as additional state variables for joint estimation. This is because there may be lack of accurate controller information or the control models may have poor quality. Therefore, it can be concluded that only the filtering step needs to be coded along with the playback function for practical DSE implementation. It should be noted that we do not need to wait for an event to trigger the DSE. As long as the measured vector $\bm{u}_{k}$, typically the voltage magnitudes and angles/frequency, is fed into the playback function, the DSE can be implemented to continuously monitor dynamic states. Note that the observability and required measurements should be provided when implementing DSE \cite{Abur_observability2017}.

\section{Roles of DSE in Power Systems Modeling, Monitoring and Operation}
Three major application areas that can potentially benefit from DSE have been identified and elaborated in this paper, namely: a) modeling, b) monitoring via the enhanced dynamic visibility, and c) operation. Each of them is discussed below.

\vspace{-0.3cm}
\subsection{Modeling}
\begin{figure}[htb!]
\centering
\includegraphics[width=\linewidth]{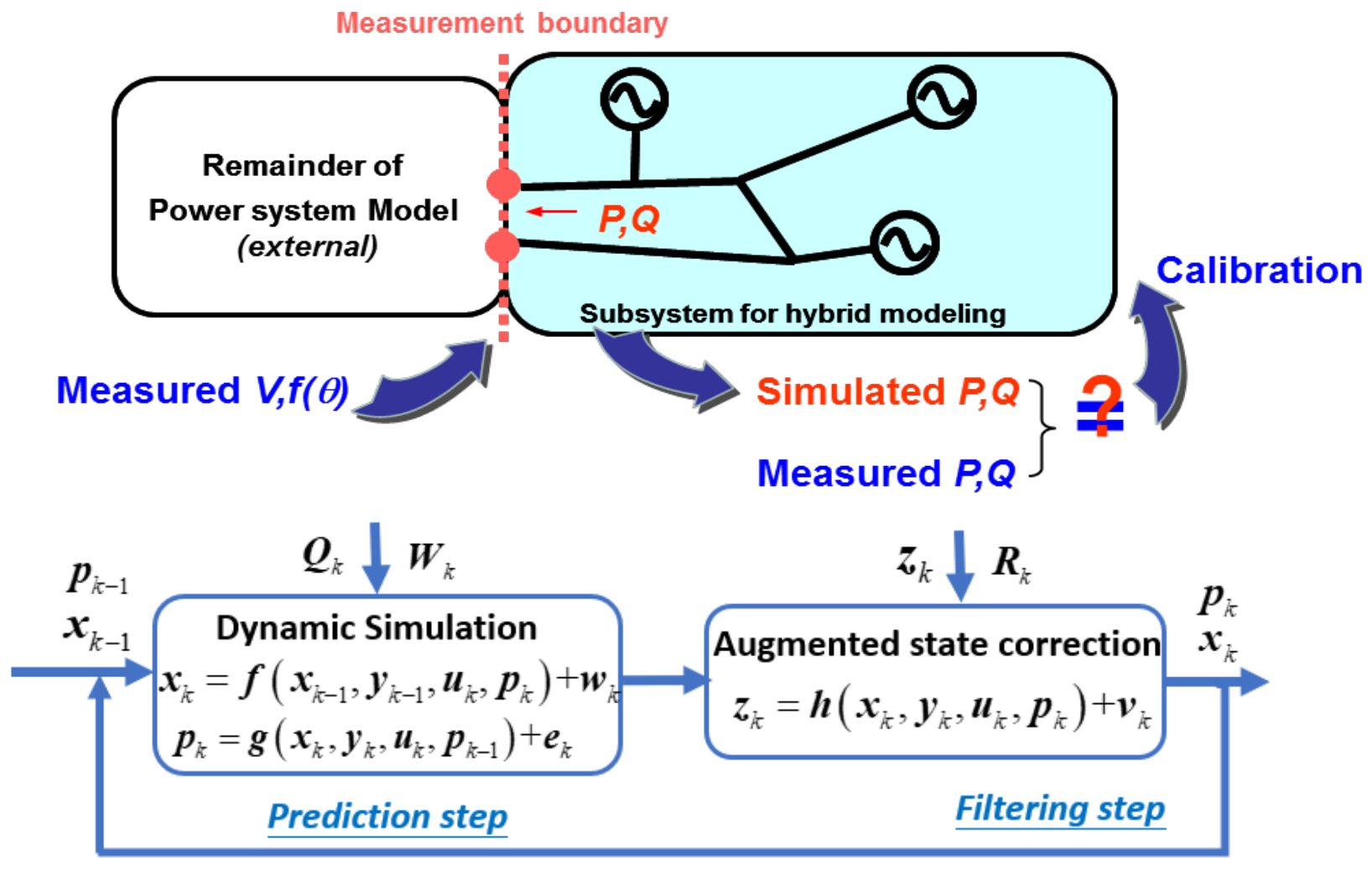}
\caption{DSE-enabled model validation and calibration.}
\label{fig4}
\end{figure}

Following the 2003 Northeastern U.S. and Eastern Canada blackout, the North America Electric Reliability Corporation (NERC) has been developing standards for periodical model validation \cite{NERC_validation_2016}. In the existing commercial software, the ``event playback'' function is leveraged to validate dynamic models using PMU data. The key idea is to take the measured generator terminal voltage magnitude/phase angle or frequency as model inputs to obtain model outputs of real power $P$ and reactive power $Q$. These responses are compared with the measured $P$ and $Q$ to validate the model adequacy \cite{Henry_calibration2013,Henry_calibration2018, Sharokh_calibration2020}. When the model of the subsystem is accurate, the simulation responses should match statistically well the actual responses. By contrast, when significant mismatches are observed, the model is considered inadequate and parameter calibration is needed. This process can be found in the upper part of Fig. \ref{fig4}, where $\bm{e}_{k}$ is the parameter error with covariance matrix $\bm{W}_{k}$ and $\bm{g}$ is the parameter regression function. The difference between Figs. \ref{fig3} and \ref{fig4} is that additional parameters are augmented with the states for joint estimation. Note that the criterion for model validation is a byproduct of DSE, namely the innovation vector-based statistical test. Furthermore, the targeted dynamic components can be synchronous machines, wind farms, power electronics-interfaced DERs \cite{YC_DERS_2015} that can be described by DAEs.

To calibrate the system models, both the system model structure and parameters need to be carefully investigated. Traditionally, they can be found from manufacturer databases. However, parameters might drift during operations due to a variety of factors, e.g., environmental changes, the aging process, and coupling effects. Taking generators offline for testing and parameter calibration is costly and affects the system reliability. Therefore, online calibration has been an important application of DSE. Since not all parameters are identified using the present measurement sets, it is, therefore, critical to select appropriate candidate parameters before starting a parameter calibration procedure. The calibration typically includes four main steps \cite{Henry_calibration2013,Henry_calibration2018}, namely 1) initial checks, 2) sensitivity analysis, 3) parameter estimation and 4) parameter validation using other disturbances. The first step is to eliminate obvious errors in parameters that are not realistic, such as fractional values for the model flags that should be integers, parameter values outside the normal ranges, swapped values for limiters, and incorrect statuses of controllers. The sensitivity analysis provides a guideline in determining parameter sensitivity for different disturbances and identifying the candidate parameters for calibration. The selected parameter vector $\bm{p}$ is augmented with the original state vector for joint estimation as shown in the lower part of Fig. \ref{fig4}. The joint DSE methods are based on, e.g., EKF, UKF, constrained UKF, and EnKF \cite{Abur_constrainedUKF_2018,Valverde2011,  Rui_EnKF2015,Exposito_2019}. After calibration, it is necessary to perform model validation using different events so that the identified parameters are not local optima. If the model deficiency for other events is detected, the calibration process should be repeated until there are no inconsistent responses between models and measurements. Besides the model validation and calibration, the power system coherency identification would also benefit from the estimated dynamic states, especially the rotor angles and speeds \cite{Joe_coherency2004,  Aghamohammadi_coherency2016}. The identified coherency allows the development of model reduction for stability assessment and islanding control. Another potential application of DSE is the dynamic reduction of large power systems \cite{Chakrabortty2011}, where the equivalent model parameters can be estimated together with the machine dynamic states. The dynamic load parameter identification and the unknown machine parameters, such as inertia, sub-transient reactance, governor dead-band, etc., can also be cast into the state-space formula. Then, various DSE algorithms can be utilized for unknown parameter estimation \cite{Abur_parameter_estimation2016,Pal2015,Pal2018_deadband}.

\vspace{-0.3cm}
\subsection{Monitoring}
\begin{figure}[htb!]
\centering
\includegraphics[width=\linewidth]{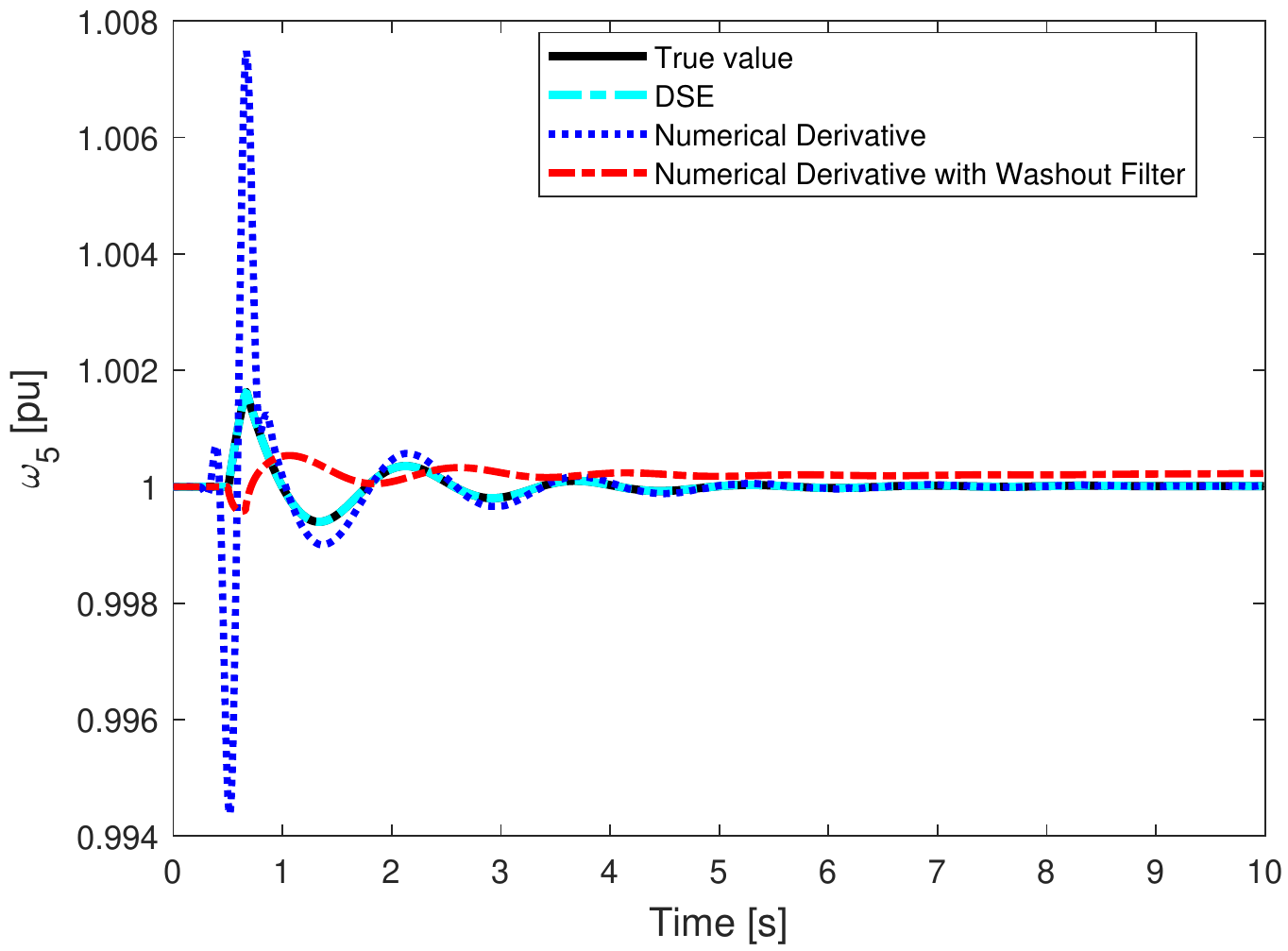}
\caption{Comparisons of different methods for internal generator frequency estimation.}
\label{fig5}
\end{figure}

One of the key DSE applications is to provide enhanced dynamic visibility for power system monitoring. Part of the dynamic visibility of DSE has been shown in Fig. \ref{fig2} but a more detailed summary includes seven key functions: i) dynamic state trajectory tracking, ii) oscillation monitoring, iii) bus frequency, ROCOF and center of inertia (COI) frequency estimation, iv) data quality detection and correction, including cyber attacks, v) unknown control inputs identification, vi) anomaly detection and vii) other applications that involve with larger time constants than the electromechanical dynamics. The dynamic state trajectory tracking is a natural product of DSE as it provides the time series of the generator and its controller states in the presence of system disturbance. Among these, the generator rotor speed and angle have been widely used in power system, such as oscillation detection \cite{Sanchez2012}. For example, in damping-torque-based forced oscillation source location, the rotor speed, and angle are required \cite{Maslennikov2016}. The estimated machine rotor speeds also allow us to resort to the frequency divider for bus frequency estimation \cite{Junbo_frequency_2018}. This interesting result reveals that if the system machine rotor speeds are available, all bus frequencies can be estimated. Since the number of machines is much smaller than the number of buses, it significantly reduces the requirement of PMU installations for bus frequency monitoring. The real-time COI frequency plays an important role in power system stability analysis and control. Via the DSE outputs, we can obtain the COI frequency at the same refreshing rate of the PMU measurements. This provides the online reference frequency for control \cite{Junbo_COI_2019}. Note that a widely used industry practice is to leverage generator terminal voltage and current measurements to calculate an approximated internal machine rotor speed. However, its accuracy is questionable as can be seen in the following example: a three-phase short-circuit occurs at Bus 16 of the IEEE 39-bus system on $t$=0.5s and the fault is cleared after one cycle by opening the transmission line connecting buses 16 and 17. The machines are modeled using the subtransient model with a DC1A exciter, PSS1A stabilizer, and IEEE-G3 governor. Fig. \ref{fig5} shows the estimated generator 5 internal rotor speed by DSE, numerical derivative with low pass filter and numerical derivative with washout filter \cite{Federico_2017} of the derived generator angle from the terminal voltage and current measurements. As expected, the numerical derivative of the approximated rotor angle suffers from numerical errors at the moment the event happens; numerical derivative with washout filter is able to address that but at the cost of modifying the time-domain response. By contrast, the DSE can accurately track the rotor speed. On the other hand, ROCoF has been utilized as an important parameter for system protection, especially for the inverter-based generations. By treating it as a dynamic state instead of an algebraic variable, an accurate estimation of it can be achieved \cite{Abhinav2019}.

The measurements are always subject to noise and even large errors caused by communications, instrument transformer saturation, cyber-attacks, etc. It is important that DSE naturally filters out the measurement noise and provides more accurate data for other applications. Since the field measurements typically follow non-Gaussian distributions \cite{Junbo_non_Gaussian2017}, robust DSE developed based on the robust statistics theory is needed \cite{Junbo_constrained_UKF2019}. While for randomly occurred bad measurements, both normalized residual statistical tests and robust detection are developed. However, it is shown that the threshold for the traditional normalized residual statistical test is not analytical and depends on different systems characteristics. Also, it does not provide reliable outputs in the presence of unknown measurement noise statistics. This is not the case for the robust estimation that automatically detects and suppresses bad data. For the cyber-attack scenario, robust DSE with high breakdown points or novel machine learning-aided DSE might offer a solution \cite{Yachine2020}. This is still an open area that requires further investigation.

Another important visibility is on the excitation system that could significantly affect the system stability. Although PMU measurements might be used to calculate the excitation voltage in certain cases \cite{Tianshu2007}, this cannot be generalized to the modern brushless excitation systems. Furthermore, under stressed conditions, the excitation voltage is restricted by timer-based over-excitation limiters, which may dramatically affect the system stability margin \cite{Anagnostou_excitor_2016}.
Traditional DSA model typically does not capture if the state variables corresponding to exciter output are saturated or not. When saturation occurs, it is no longer a state variable and thus should be kept as an unknown input estimation. To achieve this goal, the excitation voltage is taken as the unknown variable for a two-stage estimation, where the estimation of input is done first and then substituted into the original model for DSE. Both EKF \cite{Kamwa2011_unknown_Inputs, Ghahremani2016_unknown} and UKF \cite{Anagnostou2017_unknown_inputs2017} considering unknown excitation voltage are used for that and the UKF has a better capability of handling the model nonlinearities and saturation effect. These two approaches require the local generator frequency to ensure the observability of unknown inputs. However, the local generator frequency is typically not measured directly by the PMUs. To deal with that, a correlation aided robust DSE for unknown input and state estimation is proposed in \cite{Junbo_unknown_inputs2019}. Compared with the previous works, no generator frequency measurement is required, and it has better robustness in dealing with bad data. It is worth noting that the unknown (i.e., not measured) mechanical torque of the turbine-governor system is also estimated using the approach in \cite{Junbo_unknown_inputs2019}. 

Besides the unknown control inputs, there are also anomalies, such as controller failures or malfunctions that can affect the system stability. For example, if the over-excitation occurs, a voltage security issue may arise. Furthermore, if the failure of the excitation system happens but without being detected, the accuracy of the differential-algebraic equations will be negatively impacted, leading to incorrect conclusions about system stability. It is very important to detect these failures timely to avoid the risk of exciter and voltage regulator damages. By relying on the estimated dynamic states and checking the consistency between the control model outputs and expected outputs, the over-excitation and abnormal mechanical power changes can be detected \cite{Marchi2020}. The multiple models based DSE technique is also developed to detect excitation failures in \cite{Abur_excitor_2016}. Those techniques may be further extended to deal with the anomalies of other controllers, such as governor, and power system stabilizer.

Besides monitoring electromechanical dynamics, there are other DSE-based monitoring applications that use DAEs but with much higher time constants. This is the case, for instance, of the increasingly important real-time thermal rating of lines and cables, which is drastically changing the customary way in which static security assessment is performed. Indeed, the existing assets are reaching their conservative ampacity limits, typically defined on a seasonal basis, without due consideration to meteorological conditions (temperature, wind speed, etc.). The simultaneous, real-time estimation of the external parameters arising in the thermal models of lines and cables can help us improve the operation of underground cables \cite{FLeon2015,Exposito2018_smartcity}.

\begin{figure}[htb!]
\centering
\includegraphics[width=\linewidth]{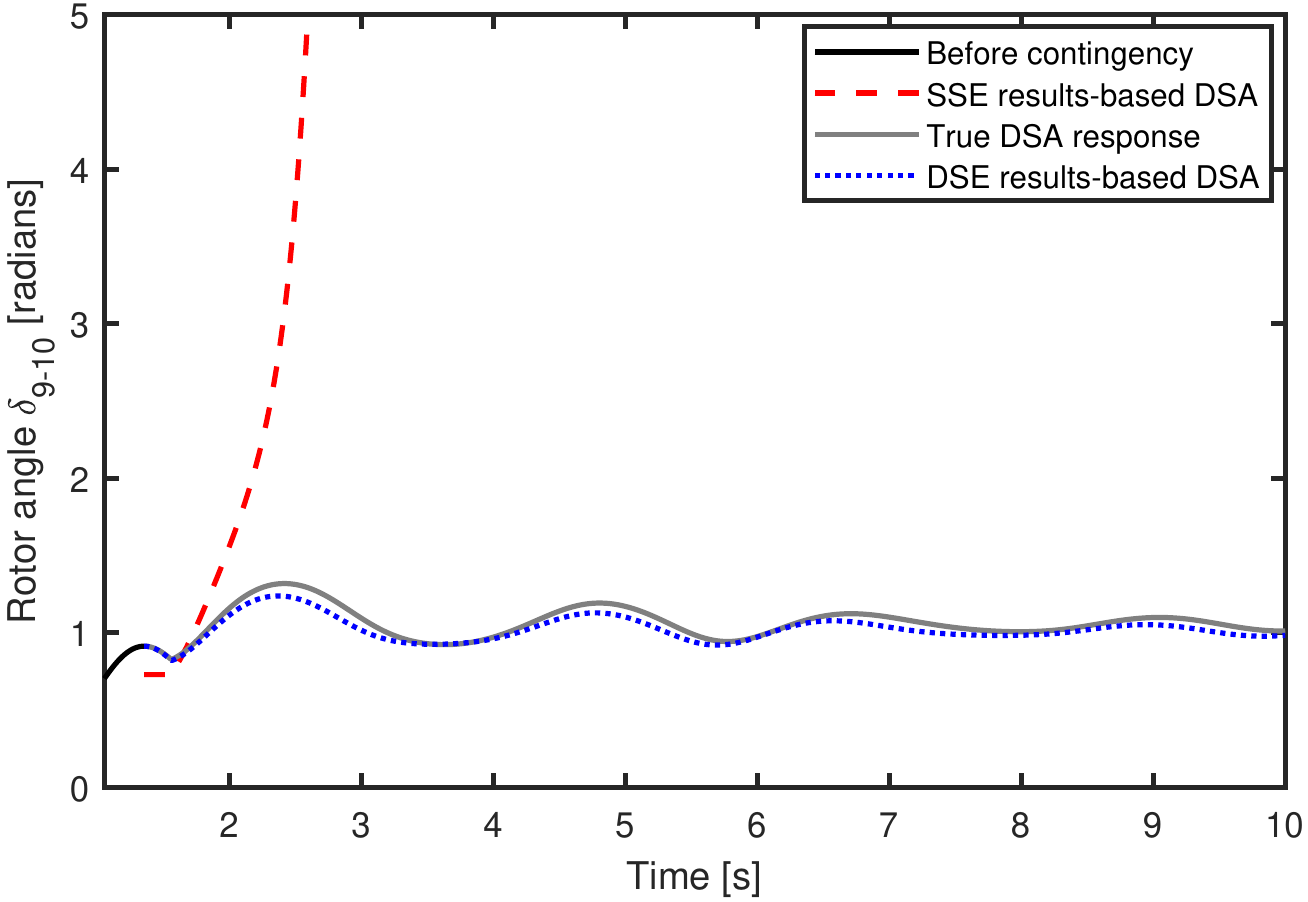}
\caption{Influence of initial conditions on DSA results.}
\label{fig6}
\end{figure}

\vspace{-0.4cm}
\subsection{Operation}
SSE plays an important role in today's power system operations, especially the essential data base for contingency analysis, static voltage stability and optimal power flow. Compared with SSE, DSE outputs would also offer new benefits to today's EMS. Two main categories are elaborated here, namely DSA and dynamic stability assessment.   

Model-based DSA relies on the initialization of dynamic models. The inaccurate initialization of variables might lead to different conclusions about whether the system is stable or not. In the legacy power systems, the operating point evolves slowly in time and the SSE results can be used to calculate the system's initial conditions that are required by DSA tools \cite{Leiwang_DSA_2006}. In other words, if the system is in equilibrium at 60Hz (or 50Hz), SSE is adequate for initializing the differential equations, which is how DSA is done in today's EMS. When the system is not at equilibrium or not at 60Hz (or 50Hz), SSE can’t be used to initialize the differential equations. Indeed, in future power systems with high penetration of DERs, the variables will change more often and more rapidly, rendering SSE too slow to provide DSA tools with an accurate picture of the system states. But it is during this time period that DSA would be even more important for power system operation. One example is the 2003 Northeast blackout, where the system frequency deviated from 60Hz for hours and DSA would help to guide operator's restorative actions. Model initialization for DSA is therefore the first important role of DSE. This is demonstrated through a simple numerical simulation carried out on the IEEE 39-bus system. The system is assumed to be operated under the scenario, where the loads and DERs have large stochastic behaviors, yielding some oscillations of the synchronous generators. The operator would like to assess if the system is able to withstand a contingency. To this end,   
at $t$=1.3s, a three-phase short-circuit is applied at Bus 28 and cleared after 30ms by opening the transmission line connecting Buses 28 and 29. Two cases are considered: 1) the model is initialized by using the estimated bus voltage magnitudes and angles obtained from SSE at $t$=1.3s; 2) the model is initialized by using the dynamic states obtained from DSE at $t$=1.3s. DSA is then performed in both cases. The rotor angle of Generator 9 with respect to that of Generator 10 is displayed in Fig. \ref{fig6}. It can be observed that the SSE-based DSA indicates that the system loses stability while the actual system remains stable. By contrast, the DSE-based DSA reflects true system behavior. The key insight is that in the SSE-based initialization, the derivatives of dynamic state variables are set to zeros based on the quasi-steady-state assumption, but the actual system already deviates from a steady state due to the variations of relative large loads and DERs. As a result, the calculated states using SSE-based DSA are not accurate. This is not the case for DSE-based DSA as their results contain the non-zero derivatives and can be directly used for non quasi-steady-state initialization.
\begin{figure}[htb!]
	\centering
	\includegraphics[width=\linewidth]{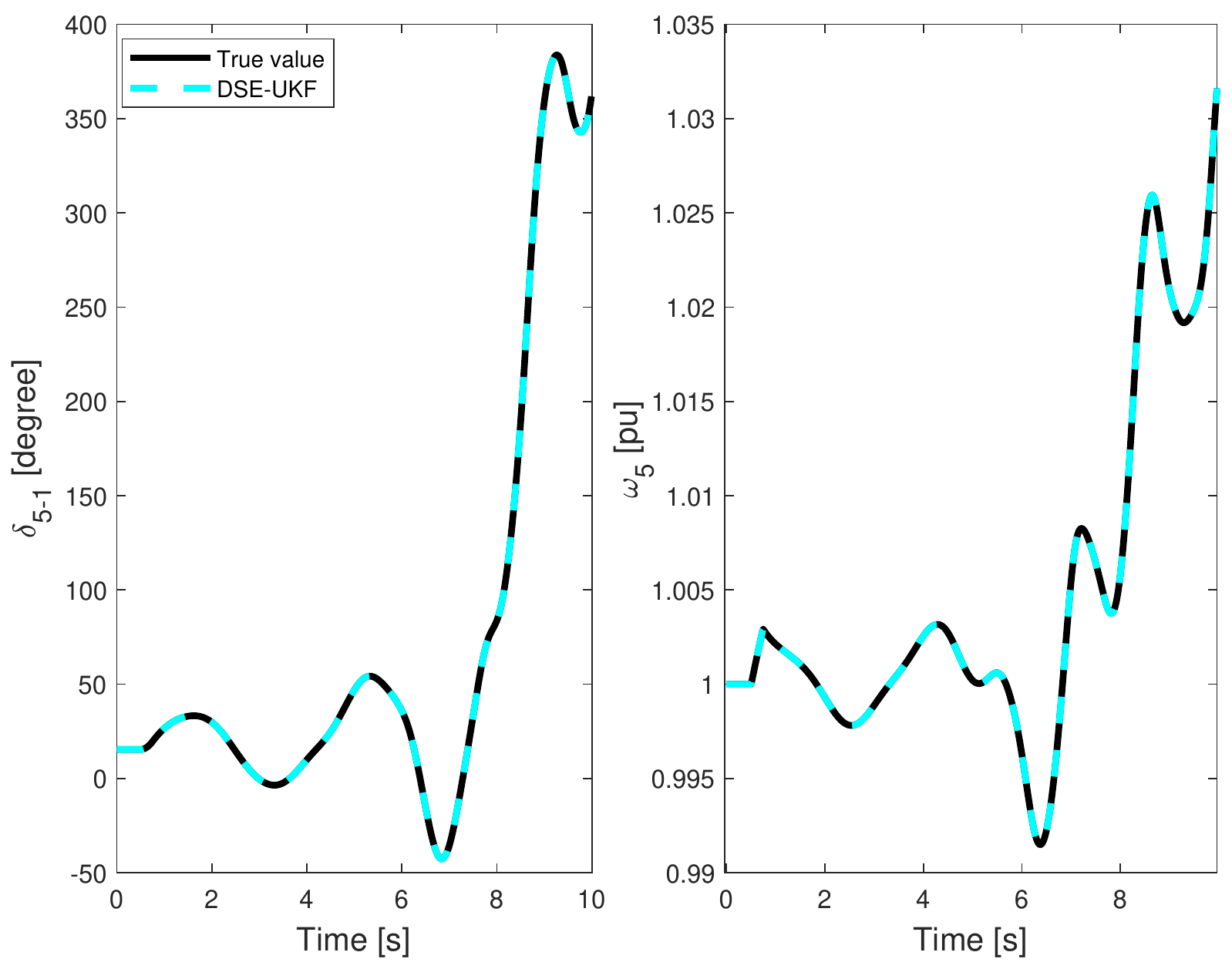}
	\caption{DSE for unstable system with high non-linearity.}
	\label{fig7}
\end{figure}

Dynamic stability assessment typically relies on a database that is built from offline simulations, wherein the evolution of the system dynamic states following different events is recorded. Then, an online classifier algorithm, e.g. decision tree or random forest, continuously compares real-time measurements with the database. If an evolving instability is detected, remedial control actions are initiated. This control philosophy is deployed in many power systems around the world. Due to its dependence on offline simulations, it is challenging to consider all possible operating scenarios during offline simulations \cite{Ortega2018}. This is particularly important with the increasing deployment of power electronics-interfaced DERs. DSE offers an alternative to these databases as it can provide a picture of the system dynamics in almost real-time, and can be used to compute transient stability indices, e.g., rotor angle stability \cite{Kamwa2019_stability}. This alternative control philosophy based on real-time information is yet in its infancy but seems to be promising. Rotor angle stability can also be assessed via the sign of the system’s maximal Lyapunov exponent \cite{Wei_rotorstab_2018}, where the generator rotor angles provided by DSE would be of importance to predict system stability. To further demonstrate the capability of DSE in dealing with system instability, the same three-phase fault for the results in Fig. \ref{fig5} is used but with 30 cycles of clearing time. This leads to system instability as shown in Fig. \ref{fig7}. The robust UKF-based DSE \cite{Junbo_GMUKF2019} is utilized to estimate and track generator rotor angle and speed.
We observe that even on the verge of losing stability, where the system response is highly nonlinear, DSE continues to track the evolution of the system states that can be used for emergency/ in extremis control schemes. On the other hand, the status and/or states of dynamic devices are playing a key role in dynamic voltage stability, such as overexcitation limiters, and dynamic loads. DSE can be appropriately formulated to monitor them and provide insightful information on voltage stability conditions of the system. For future power electronics-dominated systems, the frequency issue is of critical concern due to the reduced system inertia. DSE can provide online frequency information to enable more effective system frequency stability assessment. However, there are very limited researches along these directions and more investigations are needed.

\vspace{-0.3cm}
\subsection{DSE for Power Electronics-Interfaced Renewable Generation Visibility}
Renewable energy sources (RESs) are typically integrated with the grid through power electronics converters. To achieve reliable and cost-effective RESs integration, dynamic visibility is needed. Compared with the conventional synchronous generators, RESs exhibit a higher level of variability and uncertainty. Furthermore, the power converters and the control loops of RESs are very different from the synchronous machines. The power electronics have much smaller time constants than those of the synchronous generators, yielding much faster dynamics. As a result, the measurement system used for the monitoring of RESs should have a faster sampling speed and broader bandwidths for data transmission. It should be noted that besides the traditional voltage stability, rotor angle stability and frequency stability, there is also converter-induced stability \cite{stability_definition_2020}. For monitoring and control such fast dynamics of the converters, DSE would be a good choice. For example, the UKF-based DSE is developed to extract the fundamental components of the point of common coupling voltage and load current for grid synchronization of PV systems in \cite{Swain2018}. The doubly fed induction generator (DFIG) wind generator's model can be put into the state-space form with their dynamic states estimated by DSE to achieve visibility. The DSE can be implemented via UKF, particle filter, and unscented particle filter \cite{Yu2016_DFIG,Yu2018_DFIG,YU2019_PF_DFIG}.
Since there are errors with the wind speed measurements, DSE for DFIG with unknown wind speeds is addressed in \cite{Anagnostou2019_DFIG}. The joint estimation of dynamic states and parameters of permanent-magnet synchronous motor-based wind generator is investigated in \cite{Exposito2020_PMSM}. It should be noted that the relation of estimated states via DSE with system dynamics and stability phenomena is still the subject of ongoing research. Except for the visibility of dynamic states, it is also important to detect any anomalies inside the RESs, such as control failures, erroneous tripping actions, etc. This would provide timely information for operators to take proactive controls for maintaining system stability. For example, due to the erroneous frequency and ROCOF measurements provided by the phase-locked-loop, several large solar farms have been incorrectly tripped in the 2016 California blue cut fire event \cite{Bule_cut2017}. The DSE would allow us to provide more accurate frequency and ROCOF estimates and visibility of abnormal behaviors by checking the model and measurement consistency. This is an open problem that needs more investigations.

\vspace{-0.3cm}
\section{Conclusions and Future Work}
This paper provides a comprehensive summary of the roles of DSE for power system modeling, monitoring and operation. The relationships between SSE and DSE have been identified from the implementation requirements and related EMS functionalities. Several representative examples have been presented to highlight the critical importance of DSE for future power systems, especially for the development of the next-generation EMS. 

Future research on DSE can be generally categorized into four key aspects: data infrastructure for DSE, DSE core functions, DSE applications, and its practical implementation to support the operation and planning of future power systems.
\begin{itemize}
\item \textbf{Data Infrastructure for DSE:} DSE, whether being implemented in a centralized or distributed/decentralized manner, needs real-time measurements, and these measurements need to be transferred to proper locations, such as a control center, a generation facility, or a substation. The power industry has been moving forward in the deployment of both sensors and communication networks. Research topics include: what are the data requirements such as data rates, signals, sensor placement? What are the data communication requirements, such as bandwidth, reliability, and redundancy for wide-area system applications? How to achieve more efficient and robust parallel and/or distributed implementations of DSE? And for a hierarchical system, how would the data communication network be structured and where data should be sent to for the DSE application?
\item \textbf{DSE Core Functions:} With the active efforts by many researchers for more than a decade, many DSE algorithms have been developed, and the DSE performance has been significantly improved, as summarized in this paper. However, power systems are evolving to be more complex and different, so DSE core functions should continue to improve to meet the operational requirements of future power systems. Research topics include: How to improve computational aspects of DSE to meet real-time requirements for large-scale systems? How to formulate DSE with fast inverter dynamics and potentially deal with mixed slow and fast dynamics? Can machine learning or other data analytical approaches help improve DSE performance in the context of a large amount of data and a large number of system configurations? And how to break DSE into pieces in real-time for islanding situations, especially for a more resilient power system?
\item \textbf{Development of DSE Applications:} DSE provides unprecedented detailed dynamic information, compared with SSE. A large number of applications in modeling, monitoring and operation can benefit from DSE as mentioned in this paper. Many more are emerging. There are conventional applications that can be enhanced by DSE-provided dynamic information and new applications that can be enabled by DSE. Some of them include: aggregated model calibration for wind/solar farms, loads and DERs; oscillation source location; look-ahead DSA; visibility and detection of converter-induced instability; anomaly detection of DERs, where the complicated control loop may have fault or failures etc.
\item \textbf{Practicality of DSE Applications:} All the research efforts have a common goal which is to make DSE a practical part of the power system functions. Model calibration is an example of DSE applications, which is already in commercial tools, reliability standards, and industry practices. For most DSE applications, there are still significant gaps to address in terms of their practicality. Research questions include: How to transition from SSE to DSE while SSE and DSE will most likely co-exist for a significant period? How to make DSE compatibility with the control room environment in terms of its information technology infrastructure and human-machine interface? For example, DSE results are updated too fast for operators to take actions and how to present the critical information only to operators would be an important research need. This may be addressed by development of advanced AI tools fed with DSE data streaming for visibility and stability assessment; What training should be developed and provided to prepare the workforce for DSE applications? And what new standards are needed to enable the DSE applications?
\end{itemize}

\ifCLASSOPTIONcaptionsoff
\newpage
\fi

%\bibliographystyle{IEEEtran}
%\bibliography{lib}

\end{document}